\documentclass[pra,twocolumn,showpacs,tightenlines,nofootinbib,superscriptaddress]{revtex4}

\newcommand{\ket}[1]{\left|#1\right\rangle}

\newcommand{\sandwich}[3]{\left \langle #1 | #2|#3\right\rangle}

\usepackage[utf8x]{inputenc}
\usepackage{graphicx}
\usepackage{dcolumn}
\usepackage{bm}
\usepackage{amsmath}
\usepackage{amssymb}
\usepackage{subfigure}
\usepackage{textcase}
\usepackage{hyperref}

\begin{document}

\title{A theoretical study of the $g$-factor and lifetime of the $6s6p\, ^3\!P_0$ state of mercury}

\author{S.~G.~Porsev}
\affiliation{Department of Physics and Astronomy, University of Delaware, Newark, DE 19716, USA}
\affiliation{Petersburg Nuclear Physics Institute, Gatchina, Leningrad District 188300, Russia}

\author{U.~I.~Safronova}
\affiliation{Department of Physics, University of Nevada, Reno, Nevada 89557, USA}

\author{M.~S.~Safronova}
\affiliation{Department of Physics and Astronomy, University of Delaware, Newark, DE 19716, USA}
\affiliation{Joint Quantum Institute, National Institute of Standards and Technology and the University of Maryland,
College Park, Maryland, 20742, USA}

\date{\today}

\begin{abstract}

We calculate the $g$-factor of the $6s6p\, ^3\!P_0$ state of 199 and 201 mercury isotopes using a relativistic
high-precision all-order method that combines the configuration interaction and the coupled-cluster approaches.
Our values $g(^{199}{\rm Hg}) = -0.9485(49) \times 10^{-3}$  and $g(^{201}{\rm Hg}) = -0.3504(18) \times 10^{-3}$ are in agreement
with the experimental measurements within the 0.5\% theoretical uncertainty. We also calculate the hyperfine quenching rate of the $6s6p\,\, ^3\!P_0$ state in $^{199}$Hg and $^{201}$Hg and determine its lifetime to be 1.3 and 1.9~s, respectively.
\end{abstract}


\maketitle

\section{Introduction}

Mercury is one of a most promising candidates for a frequency standard due to its low susceptibility to the blackbody radiation (BBR). Recently, an optical lattice clock based on the clock transition between the $^1\!S_0$ and $^3\!P_0$ states of $^{199}$Hg has achieved the uncertainty at the level of $8 \times 10^{-17}$, with the dominating uncertainty being from a frequency shift introduced by the trapping lattice light \cite{YamOhmUsh15}. To reduce this uncertainty further, thorough understanding of the mercury atomic structure is crucial. This work provides a benchmark test of theoretical accuracy in evaluating Hg atomic properties, an important step towards understanding the Hg complicated  electronic structure. Accurate theoretical calculations of Hg properties, including the BBR and various sources of AC Stark shifts will be needed to further reduce the uncertainty of Hg atomic clock.
For such calculations reliable estimates of the theoretical uncertainties are required.

An atomic response to externally applied magnetic field is characterized by a dimensionless number called the $g$-factor. For the $^1\!S_0$ ground state of $^{199}$Hg, the $g$-factor was measured with a relative uncertainty at the $8\times 10^{-7}$ level~\cite{Cag61,AfaBakBan14}. For the $^3\!P_0$ state of $^{199}$Hg, which is of our interest because of its importance for Hg frequency standard, the $g$-factor was calculated and measured in Ref.~\cite{LahMar75}.
The authors used a semi-empirical method of calculation and estimated the theoretical accuracy at the level of 1\%, while
the experimental precision was considerably greater.

A precise knowledge of the $g$-factor of the clock states is crucial in suppressing frequency shifts from magnetic field to the level of $10^{-19}$ in a recent microwave-dressing scheme proposed by Zanon-Willette \textit{et. al.} \cite{ZanCleAri12}.
Since the calculation of the $g$-factor involves the same matrix elements as those used in the calculation of the lifetime of the $^3\!P_0$ state of $^{199}$Hg, it is important to revisit the problem of an accurate and reliable calculation of this quantity.

We carried out the calculation of the $g$-factor of the $6s6p\,\,^3\!P_0$ state in the framework of a pure \textit{ab initio} relativistic hybrid method combining the configuration interaction (CI) with the single-double coupled cluster approach [CI+all-order method, see Refs.~\cite{Koz04,SafKozJoh09}] and found an excellent agreement with available experimental results. We have also calculated hyperfine structure (HFS) constants
of the low-lying odd-parity states and the lifetime of the $^3\!P_0$ state for $^{199}$Hg, resolving the discrepancy of theory and experiment for the latter.

\section{Calculation of the $6s6p \, ^3\!P_0$ $g$-factor}
\subsection{Formulation of the problem}
If an atom is placed in an external magnetic field $\mathbf{B}$, the interaction of the atomic magnetic moment $\bm \mu$ with $\mathbf{B}$ is described by the Hamiltonian
\begin{equation}
H = - {\boldsymbol \mu} \cdot \mathbf{B}.
\label{eq:magnetic_H}
\end{equation}
The atomic magnetic moment $\bm \mu$ is a sum of the electronic magnetic moment $\bm \mu_e$ and the nuclear magnetic moment $\bm \mu_I$.

We describe an atomic state $\ket{\psi} = \ket{F M_F}$ in the basis of its total angular
momentum $\mathbf{F = J+I}$, where $\mathbf{J}$ and $\mathbf{I}$ are the total electronic and nuclear angular momenta, respectively, and $M_F$ is the magnetic quantum number.

Using Eq.~(\ref{eq:magnetic_H}), assuming that ${\bm \mu} = -\mu_0 g {\bf F}$, and
directing the external magnetic field $\mathbf{B}$ along the $z$-axis,
we easily obtain for the energy shift of the atomic state $\ket{\psi}$:
\begin{equation}
\Delta E = \sandwich{\psi}{H}{\psi} = g \mu_0 M_F B,
\label{eq:g_energy}
\end{equation}
where $\mu_0$ is the Bohr magneton and $g$ is a dimensionless $g$-factor.

We focus our analysis on the odd $^{199}$Hg and $^{201}$Hg isotopes of mercury with the 1/2 and 3/2
nuclear spin $I$, respectively. Due to the non-zero nuclear spin, the atomic state $\ket{\psi}$ differs from the bare atomic state owing to the hyperfine structure interaction $H_\text{hfs}$.
The resulting $^3\!P_0$ state wave function, including the first-order correction, is given by
\begin{eqnarray}
\ket{\psi} = \ket{^3\!{P}_0, I M_F}+\sum_n
\frac{|n\rangle \langle n |{H_\text{hfs}}| {^3\!P_0, I M_F}\rangle}{E_{^3\!{P}_0}-E_n},
\end{eqnarray}
where $\ket{n}$ denotes bare atomic wave functions and $F=I$.

Substituting this wave function into Eq.~(\ref{eq:g_energy}), we have
\begin{eqnarray}
&& g \mu_0 M_F B \approx \sandwich{^3\!{P}_0, I M_F}{H}{^3\!{P}_0, I M_F}  \nonumber \\
&+& 2 \sum_n\frac{\sandwich{^3\!{P}_0, I M_F}{H}{n}\sandwich{n}{H_\text{hfs}}{^3\!{P}_0, I M_F}}
{E_{^3\!{P}_0}-E_n},
 \label{eq:001}
\end{eqnarray}
where we keep the terms up to the first order in $H_\text{hfs}$ and use the fact that
both $H$ and $H_\text{hfs}$ operators are real. Therefore, the calculation of the $g$-factor involves evaluation of the matrix elements of two operators: $H$ and $H_\text{hfs}$.
\subsection{Matrix elements of the $H_\text{hfs}$ operator}
The hyperfine structure coupling due to nuclear multipole moments may be represented as a scalar product of
two tensors of rank $k$,
\begin{equation*}
H_{\rm hfs} = \sum_k \mathbf{N}^{(k)}\cdot \mathbf{T}^{(k)} ,
\end{equation*}
where $\mathbf{N}^{(k)}$ and $\mathbf{T}^{\left( k\right) }$ act in the space
of nuclear and electronic coordinates, respectively. Using this expression we
 write the matrix element for $H_{\mathrm{hfs}}$ as
\begin{eqnarray*}
&&\langle \gamma' IJ'; F' M'_F |H_{\mathrm{hfs}}|\gamma IJ; F M_F \rangle = (-1)^{I+J'+F}  \\
&\times& \sum_k \langle I ||N^{(k)}|| I \rangle \langle \gamma' J' ||T^{(k)}||\gamma J\rangle
\left\{
\begin{tabular}{lll}
$I$ & $I$  & $k$ \\
$J$ & $J'$ & $F$
\end{tabular}
\right\} \\
&\times& \delta_{F'F} \delta_{M'_F M_F} ,
\end{eqnarray*}
where $J$ is the total angular momentum of the electrons and
$\gamma$ encapsulates all other electronic quantum numbers. For this calculation, we restrict the treatment of  $H_{\mathrm{hfs}}$ to the first term in the sum over $k$, i.e., we consider only the interaction of magnetic dipole nuclear moment with the electrons.
Thus,
\begin{equation*}
H_{\mathrm{hfs}} \approx \mathbf{N}^{(1)}\cdot \mathbf{T}^{\left( 1\right) } .
\end{equation*}

For the nuclear component, we express the matrix elements $\langle I ||N^{(1)}||I\rangle $
through the nuclear magnetic dipole moment $\boldsymbol \mu_I$, which is defined as follows:
\begin{eqnarray*}
\mu_I &=& \langle I,M_I=I|({\boldsymbol \mu_I})_z|I,M_I=I \rangle \nonumber \\
&=& \sqrt{\frac{I}{(2I+1)(I+1)}} \, \langle I ||\mu_I||I \rangle .
\end{eqnarray*}
Defining $\mathbf{N}^{(1)}$ in a dimensionless form as
\begin{eqnarray*}
\mathbf{N}^{(1)} &=& {\boldsymbol \mu_I}/\mu_N ,
\end{eqnarray*}
where $\mu_N$ is the nuclear magneton,
we obtain the respective reduced matrix element \begin{eqnarray*}
\langle I ||N^{(1)}|| I\rangle &=&\sqrt{\frac{(2I+1)(I+1)}{I}} \frac{\mu_I}{\mu_{N}} .
\end{eqnarray*}

The operator $T_q^{(1)}$ can be presented as the sum of the one-particle operators
\begin{eqnarray*}
T_q^{(1)} = \sum_{i=1}^N \left( T_q^{(1)} \right)_i,
\end{eqnarray*}
where $N$ is the number of the electrons in the atom and
the one-particle operator $\left( T_q^{(1)} \right)_i$ is given by
\begin{eqnarray*}
\left( T_q^{(1)} \right)_i &=& - \frac{i \sqrt{2}\, |e|
\left( {\boldsymbol \alpha} \cdot \mathbf{C}_{1q}^{(0)}({\bf \hat{r}}_i) \right)}
{r_i^2} \, \mu_N .
\end{eqnarray*}
Here  $\mathbf{C}_{1q}^{(0)}$ is a normalized spherical harmonic, $r_i$ is the radial position
of the $i$-th electron, and $\boldsymbol \alpha$ is the Dirac matrix.
\subsection{Magnetic-dipole hyperfine structure constants}
Magnetic dipole HFS constant $A$ of an atomic state $|\gamma J \rangle$
is expressed via the matrix element  $\langle \gamma J||T^{(1)}||\gamma J \rangle$ of the
electronic tensor $\mathbf{T}^{(1)}$ as
\begin{eqnarray*}
A &=& \frac{\mu_I}{I\,\mu_N}\frac{\langle \gamma J ||T^{(1)}||\gamma J \rangle}{\sqrt{J(J+1)(2J+1)}} .
\end{eqnarray*}

To test the quality of our wave functions near the nucleus, we calculated the magnetic-dipole HFS constants $A$
for the low-lying odd-parity states $6s6p\,^{1,3}\!P_J$ and compared them with the experimental results.

The calculation was carried out for $^{201}$Hg ($I=3/2$ and $\mu_I/\mu_N = -0.560225$ \cite{Sto05}) in the CI+all-order approximation, including the Breit interaction.
We also calculated the random-phase approximation (RPA) and other, generally smaller, core-Brueckner ($\sigma$), structural-radiation (SR), two particle (2P), and normalization (Norm)) corrections. These corrections are described in detail in Refs.~\cite{DzuKozPor98,DzuFlaSil87}.

The results are presented in Table~\ref{hfs}.
\begin{table*}[tbp]
\caption{Magnetic dipole HFS constants $A$ (in MHz) for $^{201}$Hg
are presented. In the 2nd column, labeled as ``CI+All'', the results obtained in the
CI+all-order approximation are listed. In the 3rd to 7th columns, we present different corrections to the ``CI+All'' values.
The values in column labeled ``Total'' are obtained as (CI+All)+RPA+$\sigma$+SR+2P+Norm.
Last column gives the differences (in \%) between the CI+All and experimental results.}
\label{hfs}%
\begin{ruledtabular}
\begin{tabular}{lccccccccc}
                          & CI+All &  RPA &$\sigma$&  SR &  2P & Norm. & Total &  Experiment                  & Diff.(\%) \\\hline \\ [-0.2pc]
$A\,(6s6p\,\, ^3\!P_1)$   & -5499  & -560 &  283   & -45 & -82 &  252  & -5651 & -5454.569(3)~\cite{Koh61}    & 0.8 \\[0.3pc]
$A\,(6s6p\,\, ^3\!P_2)$   & -3391  & -393 &  168   & -18 & -14 &  147  & -3501 & -3352.0292(8)~\cite{McDLic60}& 1.15 \\[0.3pc]
$A\,(6s6p\,\, ^1\!P_1)$   &  1422  &  153 &  -69   &  30 & -0.1& -74   &  1462 &  1316~\cite{Lur66}           & 7.45
\end{tabular}
\end{ruledtabular}
\end{table*}
We find significant cancellations between the RPA and sum of the other corrections ($\sigma$+SR+2P+Norm) for the HFS constants. While the calculations of the wave functions and the RPA corrections are carried out  to all orders in our present method, the smaller core-Brueckner, two-particle, structural-radiation, and normalization corrections are treated in the second-order of the many-body perturbation theory (MBPT). This difference in their treatment results in an additional uncertainty due to a cancellation of these effects. The $A(^3\!P_1)$ and $A(^3\!P_2)$  HFS constants obtained at the CI+all-order stage of the calculation
are very close to the experimental values and we expect that the CI+all-order approximation will produce the most reliable results for the nondiagonal matrix elements  of the $T^{(1)}$ operator needed for the calculation of the $^3\!P_0$ $g$-factor.

A larger difference with the experiment for the $^1\!P_1$ HFS constant, 7.5\%, is likely caused by the presence of the core-excited
$5d^9 6s^2 6p~J=1$ states in the Hg spectrum which mix with the $5d^{10}6s6p\,\,^1\!P_1$ state. A similar problem is well known in Yb, where the $4f^{14}6s6p~^1\!P_1$ state is separated from the core excited  $4f^{13}5d 6s^2$ $J=1$ state by only 3789~cm$^{-1}$, leading to a severe, by a factor of 3.8, discrepancy of the theoretical hyperfine constant value
\cite{DzuFla11} with the experiment.  In Hg, the energy separation of the $5d^{10}6s 6p\,\,^1\!P_1$ and the nearest $5d^9 6s^2 6p$ $J=1$ state is much larger, 24744~cm$^{-1}$. While our agreement with experimental hyperfine constant is worse for the $^1\!P_1$ state than for the $^3\!P_{1,2}$ states, it is by far better than a similar calculation gives in Yb,
clearly demonstrating that the problem of admixture of the core-excited states, that is a reason for the Yb discrepancy, is less pronounced for Hg.
\subsection{The $^3\!P_0$ $g$-factor }
From Eq.~(\ref{eq:001}), the $g$-factor of the $^3\!P_0$ state can be expressed as
\begin{eqnarray}
g(^3\!P_0) & \approx & \frac{\sandwich{^3\!P_0,I M_F}{H}{^3\!P_0, I M_F}}{\mu_0 M_F B} \nonumber \\
&+& 2 \sum_n\frac{\sandwich{^3\!P_0, I M_F}{H}{n}\sandwich{n}{H_\text{hfs}}
{^3\!P_0, I M_F}}{\mu_0 M_F B(E_{^3\!P_0}-E_n)} \nonumber \\
&\equiv& \delta g_I + \delta g_\text{hfs}.
\end{eqnarray}
The first term can be approximated by \cite{PorDerFor04}:
\begin{equation}
\delta g_I \approx -\frac{m}{m_p} \frac{\mu_I}{I\,\mu_N}  ,
\label{eq:g_I}
\end{equation}
where $m$ and $m_p$ are the electron and proton masses, respectively.
For 201 isotope we find $\delta g_I \approx 0.203 \times 10^{-3}$ in a good agreement
with an accurate value $\delta g_I = 0.200183(4) \times 10^{-3}$ obtained in \cite{LahMar75}.

The second term is simplified using the Wigner-Eckart theorem:
\begin{eqnarray}
\delta g_\text{hfs} = \frac{2}{3}\, \frac{\mu_I}{I\,\mu_N}
\sum_n \frac{\sandwich{^3\!P_0}{|\mu_e|}{\gamma_n J_n}}{\mu_0}
\frac{\sandwich{\gamma_n J_n}{|T^{(1)}|}{^3\!P_0}}{E_n-E_{^3\!P_0}}.
\label{eq:g_hfs_sum}
\end{eqnarray}
If we keep only one term $|6s6p\, ^3\!P_1 \rangle$ in the summation over $|\gamma_n J_n \rangle$ and take into account that $\sandwich{^3\!P_0}{|\mu_e|}{^3\!P_1}=\sqrt{2}\mu_0$ in the $LS$-coupling approximation, we arrive at the formula for
$\delta g_\text{hfs}$ given in Ref.~\cite{PorDerFor04}.

We calculate the
non-diagonal reduced matrix elements of the $\mathbf{T}^{(1)}$ operator using the same method as for the HFS constants and
take the CI+all-order values as  final.

The sum in Eq.~(\ref{eq:g_hfs_sum}) is strongly dominated by the first, $6s6p\,\, ^3\!P_1$
intermediate state, which has the smallest energy separation with the $^3\!P_0$ state. The next contribution,
from the $6s6p\,\, ^1\!P_1$ state, is less than 1\%. The contribution of the subsequent
intermediate state, $6s7p\, ^3\!{P}_1$, is $\sim 10^{-8}$, which is negligible at the present level of
accuracy.

We calculate the $^{201}$Hg matrix elements needed for the evaluation of $\delta g_{\rm hfs}$ to be
\begin{eqnarray}
\langle ^{3}\!P_0||\mu _{e}||^{3}\!P_1\rangle  &=& 1.390\,\mu _{0}, \nonumber \\
\langle ^{3}\!P_1||T^{(1)}||^{3}\!P_{0}\rangle &=&-23206 \,\mathrm{MHz}
\approx -0.7741 \,\mathrm{cm}^{-1}, \nonumber \\
\langle ^{3}\!{P}_{0}||\mu_e||^{1}\text{P}_{1}\rangle  &=&-0.224\,\mu_0 \nonumber \\
\langle ^{1}\!{P}_{1}||T^{(1)}||^{3}\!{P}_{0}\rangle  &=&-12035 \,\mathrm{MHz}
\approx -0.4014 \,\mathrm{cm}^{-1},
\label{eq:coupling_1P1}
\end{eqnarray}%
Combining these matrix elements with the experimental energy differences from the NIST database~\cite{RalKraRea11}
we obtain
\begin{eqnarray*}
\delta g_{\rm hfs}(^3\!{P}_0) &\approx& \delta g_{\rm hfs}^{(^3\!{P}_1)} + \delta g_{\rm hfs}^{(^1\!{P}_1)} \\
&\approx& (0.1516 - 0.0014) \times 10^{-3} = 0.1502 \times 10^{-3}.
\end{eqnarray*}%

Because the  contribution of the $^3\!P_1$ state dominates, the uncertainty in our calculation of $\delta g_{\rm hfs}(^3\!{P}_0)$ is
determined by the uncertainty in the value of {\it nondiagonal} matrix element
$\langle ^{3}\!{P}_{1}||T^{(1)}||\,^{3}\!{P}_{0}\rangle$.
The magnetic-dipole HFS constants of the $^3\!{P}_1$ and $^3\!{P}_2$ states are smaller than the experimental values by 0.8\% and 1.2\%, respectively (see last column of Table~\ref{hfs}).
Based on this comparison, and taking into account that
$A(^3\!{P}_J) \sim \langle ^3\!{P}_J||T^{(1)}||\,^3\!{P}_J\rangle$, we assume that the uncertainty in  the diagonal matrix elements of the $\mathbf{T}^{(1)}$ operator does not exceed  1.2\%.

Our analysis shows that the $6s6p_{1/2}$ configuration contributes to the $^3\!{P}_0$ and $^3\!{P}_1$ terms at the level of 99\% and 81\%, respectively.
Therefore, we expect that the behavior of the wave functions of these two states near the nucleus should be similar. As a result, we estimate the uncertainty of the {\it nondiagonal} matrix element
$\langle ^{3}\!{P}_{1}||T^{(1)}||\,^{3}\!{P}_{0}\rangle$ to be also at the level of 1.2\%.

The hyperfine structure anomaly, which results from the variation of the magnetic dipole density distribution over the nuclear volume from nucleus to nucleus~\cite{BohWei50} is already accounted for in our estimate since any uncertainties due to this effect are already included in the difference of our values for the hyperfine constants with the experiment.

We estimated the contribution of the second-order (in  $H_{\rm hfs}$) corrections to the $g(6s6p\,^3\!{P}_0)$ to be negligible at the present level of accuracy, in
agreement with \cite{LahMar75}. Therefore, our uncertainly in the value of $\delta g_{\rm hfs}(^3\!{P}_0)$ is $\sim 1.2\%$.

Combining our result with $\delta g_{I} = 0.200183(4)\times 10^{-3}$ from~\cite{LahMar75}, we obtain for
$^{201}{\rm Hg}$:
\begin{equation}
g(6s6p\,^3\!{P}_0) = 0.3504(18) \times 10^{-3}.
\label{g201}
\end{equation}

An accurate analysis of Lahaye and Margerie~\cite{LahMar75}
gives the ratio of the $g$-factors for 201 and 199 isotopes of Hg to be
\begin{equation}
\frac{^{201}g(6s6p\,^3\!{P}_0)}{^{199}g(6s6p\,^3\!{P}_0)} = -0.369\,414(16) .
\label{g_ratio}
\end{equation}

Using Eqs.~(\ref{g201}) and (\ref{g_ratio}) we determine the $g$-factor of $^{199}{\rm Hg}$
($I=1/2$ and $\mu_I/\mu_N = 0.5058852$):
\begin{eqnarray}
g(6s6p\,^3\!{P}_0) = -0.9485(49) \times 10^{-3}.
\label{g199}
\end{eqnarray}

A comparison of our results with most accurate experimental data and theoretical values
obtained in Ref.~\cite{LahMar75} is given in Table~\ref{gfactors}. Lahaye and Margerie~\cite{LahMar75}
used a semi-empirical calculation method, following an approach of Lurio {\it et al.}~\cite{LurManNov62,Lur66}.
They expressed $g(6s6p\,^3\!{P}_0)$ through $g(6s6p\,^1\!S_0)$ and a combination of hyperfine parameters
defined in~\cite{Lur66}.
Certain parameters were found from the experimental data on the $A(^3\!P_{1,2})$ and $A(^1\!P_1)$ HFS constants.
Other parameters were assumed to be connected to each other by definite ratios.
Lahaye and Margerie carried out two calculations. In the second calculation one of ratios
between hyperfine parameters was slightly modified (see~\cite{LahMar75} for more details) what resulted in slightly different
values of the $g$-factors.
The theoretical values obtained in Ref.~\cite{LahMar75} in two these calculations are presented in the table.
The difference between them was treated by the authors as an uncertainty of the result.

In the framework of the pure {\it ab initio} method, described above, we obtained the values of the $g$-factors with the 0.5\% relative uncertainty, which is two times smaller than that in Ref.~\cite{LahMar75}.
Within the theoretical uncertainty our results are in excellent agreement with all available experimental values.
\begin{table} 
\caption{The theoretical and experimental values of $g(6s6p\,^3\!{P}_0) \times 10^3$ for $^{201}$Hg and $^{199}$Hg.
The uncertainties are given in parentheses. Two theoretical results were obtained in Ref.~\cite{LahMar75}
(see the text for details).}
\label{gfactors}%
\begin{ruledtabular}
\begin{tabular}{lll}
            &\multicolumn{1}{c}{Theory}  &\multicolumn{1}{c}{Experiment} \\
\hline \\[-0.5pc]
$^{201}$Hg  &  0.3504(18)(this work)     &  0.351058(16)~\cite{VieLah77} \\[0.1pc]
            &  0.3524~\cite{LahMar75}    &  0.35108(7)~\cite{LahMar75} \\[0.1pc]
            &  0.3486~\cite{LahMar75}    &  0.3509(4)~\cite{LahHeb70} \\[0.3pc]

$^{199}$Hg  & -0.9485(49)(this work)     & -0.950319(26)~\cite{VieLah77} \\[0.1pc]
            & -0.9541~\cite{LahMar75}    &  -0.9504(4)~\cite{LahMar75} \\[0.1pc]
            & -0.9435~\cite{LahMar75}    &  -0.9502(10)~\cite{LahHeb70}
\end{tabular}
\end{ruledtabular}
\end{table}

\section{Calculation of the lifetime of the $6s6p\,^3\!P_0$ state of $^{199}$\texorpdfstring{H\MakeLowercase{g}}{Hg} and $^{201}$\texorpdfstring{H\MakeLowercase{g}}{Hg}}

The hyperfine quenching rate of the $^3\!P_0$ state can be represented (in a.u.) by~\cite{PaeArnHaj16}:
\begin{eqnarray}
A_{\mathrm{HFS}} (^3\!P_0 \rightarrow \,^1\!S_0) =
\frac{4\, \alpha^3 \omega_0^3}{27}  \left(\frac{\mu_I}{\mu_N}\right)^2 \frac{I+1}{I} |X|^2 ,
\label{Wq}
\end{eqnarray}%
where $\alpha \approx 1/137$ is the fine-structure constant, $\omega_0 \equiv E(^3\!P_0) - E(^1\!S_0)$, and
\begin{eqnarray}
X &\equiv& \sum_n \frac{\langle ^1\!S_0 ||D||\gamma_n J_n \rangle \langle \gamma_n J_n ||T^{(1)}|| ^3\!P_0 \rangle}
{E(n)-E(^3\!P_0)} \nonumber \\
 &+& \sum_m \frac{\langle ^1\!S_0 ||T^{(1)}|| \gamma_m J_m \rangle \langle \gamma_m J_m ||D|| ^3\!P_0 \rangle}{E(m)-E(^1\!S_0)} ,
\label{Sk}
\end{eqnarray}
where ${\bf D}$ is the electric dipole moment operator.

Restricting the sum over $|\gamma_n J_n \rangle$  to the first two terms, $6s6p\,^3\!P_1$ and $6s6p\,^1\!P_1$,
and the sum over $|\gamma_m J_m \rangle$ to one term, $6s7s\,^3\!S_1$,
we obtain
\begin{widetext}
\begin{eqnarray}
&&A_{\mathrm{HFS}} (^3\!P_0 \rightarrow \, ^1\!S_0) \approx
\frac{4\, \alpha^3 \omega_0^3}{27} \left( \frac{\mu_I}{\mu_N}\right)^{2} \frac{I+1}{I}   \nonumber \\
&\times& \left\vert \frac{\langle ^1\!S_0||D||^3\!P_1\rangle
\langle ^3\!P_1||T^{(1)}||^3\!P_0\rangle }{E(^3\!P_1)-E(^3\!P_0)}
+ \frac{\langle ^1\!S_0||D||^1\!P_1\rangle \langle ^1\!P_1||T^{(1)}||^3\!P_0 \rangle}{E(^1\!P_1)-E(^3\!P_0)}
+ \frac{\langle ^1\!S_0||T^{(1)}||^3\!S_1\rangle \langle ^3\!S_1||D||^3\!P_0 \rangle}{E(^3\!S_1)-E(^1\!S_0)}\right\vert^{2}.
\label{Ahfs}
\end{eqnarray}%
\end{widetext}

The matrix elements $\langle ^1\!S_0||D||^3\!P_1\rangle = 0.4831(8)$ a.u. and $\langle ^1\!S_0||D||^1\!P_1\rangle = 2.64(3)$ a.u.,
needed for the calculation, were obtained from the experimentally known weighted average lifetimes
of the $^3\!P_1$ and  $^1\!P_1$ states, $\tau(^3\!P_1) = 118.9(4)$ ns and $\tau(^1\!P_1) = 1.34(3)$ ns~\cite{CurIrvHen01}.

Taking into account the CI+all-order matrix elements given in (\ref{eq:coupling_1P1}) and using
\begin{eqnarray*}
\langle ^1\!S_0||T^{(1)}||^3\!S_1 \rangle &=& 14610 \,\mathrm{MHz} \approx 0.4873 \,\mathrm{cm}^{-1}, \\
\langle ^3\!S_1||D||^3\!P_0\rangle &=& -1.36\,{\rm a.u.} .
\end{eqnarray*}
and the energy differences from the energy level NIST database~\cite{RalKraRea11},
we find
\begin{eqnarray*}
A_{\mathrm{HFS}} (^3\!P_0 \rightarrow \, ^1\!S_0) &\approx& \frac{4 \alpha^3 \omega_0^3}{27}
\left(\frac{\mu_I}{\mu_N}\right)^2 \frac{I+1}{I}  \nonumber \\
&\times& \left|(-2.12 -0.64 -0.11)\, \cdot10^{-4} \right|^{2}  \text{ a.u.},
\end{eqnarray*}%
where the three terms in $|...|^2$ are the contributions from the $^3\!P_1$,
$^1\!P_1$, and $^3\!S_1$ states, respectively.
Taking into account that 1 a.u. $\approx 2 \pi \times 6.5797\cdot 10^{15}$ s$^{-1}$,
we arrive at
\begin{eqnarray*}
A_{\mathrm{HFS}} (^3\!P_0 \rightarrow \, ^1\!S_0)
\approx \left(\frac{\mu_I}{\mu_N}\right)^2 \frac{I+1}{I} \times 0.99 \text{ s}^{-1}
\end{eqnarray*}%
and, finally,
\begin{equation*}
A_{\mathrm{HFS}} \left( ^{3}\!P_{0}\rightarrow \,^{1}\!S_{0}\right) \approx
\left\{
\begin{array}{l}
0.52\text{ s}^{-1}\text{, for }^{201}\text{Hg} \\
0.76\text{ s}^{-1}\text{, for }^{199}\text{Hg .}%
\end{array}%
\right.
\end{equation*}

We note that these hyperfine quenching rates are approximately 60 times greater than those published in Ref.~\cite{HacMiyPor08}, 0.013~s$^{-1}$ for $^{199}$Hg and 0.0088~s$^{-1}$ for $^{201}$Hg. Unfortunately,
Ref.~\cite{HacMiyPor08} does not contain any details on how these values were obtained or any intermediate results.
We assume that a reason of this disagreement might be a calculation error made in~\cite{HacMiyPor08}.
Our present $^{199}$Hg quenching rate, 0.76 s$^{-1}$,  is in a good agreement with the experimental value 0.692(14) s$^{-1}$~\cite{WexWilDje80}. The corresponding lifetimes of the $^3\!P_0$ state in $^{199}$Hg and $^{201}$Hg are 1.3
and 1.9~s, respectively.
\section{Discussion and summary}
Our theoretical results agree with all the experimental values within the uncertainty of the calculations,
providing excellent benchmarks of theoretical accuracy in Hg. We identified several directions towards further improvement of
Hg theory accuracy discussed below.

We find that an accurate calculation of the HFS constants requires accurate treatment of the corrections to the matrix elements of the HFS operator beyond the random-phase approximation, such as smaller core-Brueckner, structural-radiation, and normalization corrections. We treat these corrections in the second-order of MBPT, while the calculations of the wave functions and RPA corrections are done to all orders of MBPT. Since we find significant cancellations between the RPA and sum of the other corrections for the HFS constants, the difference in their treatment results in an additional uncertainty. Based on a comparison of the theoretical and experimental values of the HFS constants, we assume that if the corrections beyond RPA are included in all orders of the perturbation theory, the cancellation between different corrections will be even more pronounced.

Another improvement may stem from an inclusion of triple excitations into construction of the effective all-order Hamiltonian. The effect of the triple excitations is known to be significant for calculating HFS constants of alkali-metal atoms. All of these corrections are much smaller for the electric-dipole matrix elements and the HFS constants present excellent opportunity for benchmark testing. Both of the above method developments may be carried out by incorporating corresponding modified linearized coupled-cluster all-order codes into the CI+all-order approach.

Another effect that may affect the calculation accuracy of the Hg properties is the presence of the core-excited
$5d^9 6s^2 6p$ states in the Hg spectrum.
Our CI+all-order value of $^1\!P_1$ HFS constant differs by 7.5\% from the experiment, i.e., the agreement with experiment is worse than for the $^3\!P_{1,2}$ states.
Future improvement of the $^1\!P_1$ properties could require a consideration of the core-excited states on the same footing as the $5d^{10} nl n'l'$ states.

To summarize:

(i) Our calculation demonstrates theory ability to calculate such a complicated quantity as g-factor of the $^3\!P_0$ state with
a 0.5\% accuracy from first principles rather than semi-empirical approaches.

(ii) It provides much needed benchmark test of theoretical accuracy in the first principles evaluation of Hg atomic properties, clearly demonstrating predictive capabilities of our method for the Hg clock development.

(iii) Previous calculations of the HFS constants for the multivalent atoms generally assumed cancellation of the various corrections to the HFS operator {\it beyond} RPA. Our calculation clearly demonstrates that for Hg this is incorrect. In fact, we find that other corrections, such as the core Brueckner and structural radiation corrections, are large and nearly cancel the RPA corrections. This observation allowed us to calculate the values of the g-factors for 199 and 201 isotopes with a high accuracy. This also demonstrates a clear need for developing approaches to treat the corrections to all orders of the perturbation theory. Theoretical calculations of HFS constants are used to infer nuclear magnetic moment in systems where other methods are not available.

(iv) We have calculated the hyperfine quenching rate of the $6s6p\,\, ^3\!P_0$ state resolving the
discrepancy between the experiment and theoretical result obtained previously. An accurate knowledge of this quantity is very important for the experimentalists working with the Hg atomic clock in the $^1\!S_0 -\, ^3\!P_0$ transition. The lifetime of the excited clock state is a crucial factor in ultimate Hg clock uncertainty.

\acknowledgements
We are grateful to Thaned Pruttivarasin, Hidetoshi Katori and Noriaki Ohmae for bringing this problem to our attention, helpful discussions, and comments on the manuscript. This work was partly supported by the U.S. NSF Grant No. PHY-1404156 and  PHY-1520993.
S.P. acknowledges support from Russian Foundation for Basic Research under Grant No. 17-02-00216.


\end{document}